# A Computational Analysis of Galactic Exploration with Space Probes: Implications for the Fermi Paradox

Carlos Cotta and Álvaro Morales

ETSI Informática, Universidad de Málaga
Campus de Teatinos, 29071 Málaga - Spain

`ccottap@lcc.uma.es`

**Abstract.** Temporal explanations to the Fermi paradox state that the vast scale of the galaxy diminishes the chances of establishing contact with an extraterrestrial technological civilization (ETC) within a certain time window. This argument is tackled in this work in the context of exploration probes, whose propagation can be faster than that of a colonization wavefront. Extensive computational simulations have been done to build a numerical model of the dynamics of the exploration. A probabilistic analysis is subsequently conducted in order to obtain bounds on the number of ETCs that may be exploring the galaxy without establishing contact with Earth, depending on factors such as the number of probes they use, their lifetime and whether they leave some long-term imprint on explored systems or not. The results indicate that it is unlikely that more than $\sim 10^2$-$10^3$ ETCs are exploring the galaxy in a given Myr, if their probes have a lifetime of 50 Myr and contact evidence lasts for 1 Myr. This bound goes down to $\sim 10$ if contact evidence lasts for 100 Myr, and is also shown to be inversely proportional to the lifetime of probes. These results are interpreted in light of the Fermi paradox and are compatible with non-stationary astrobiological models in which a few ETCs have gradually appeared in the Fermi-Hart timescale.

## 1. Introduction

One of the most recurring and interesting issues regarding the search for extraterrestrial intelligence (SETI) is the so-called Fermi paradox. The absence so far of any physical evidence of advanced extraterrestrial life, in the face of the large number of potentially habitable stellar systems and the vast age of the galaxy gives rise to an apparent paradox, or –at the very least– to a problem (well-summarized in Fermi's conspicuous question "Where is everybody?") that demands an appropriate solution. To illustrate this issue, the number of systems potentially capable of harboring life in the galactic habitable zone (GHZ) [19] is $\sim 10^{10}$ [3] and the median age of Earth-like planets in the Milky Way is 1.8±0.9 Gyr greater than that of Earth [25,12]. Needless to say, there are also limiting factors to be taken into account, such as the rate at which complex life emerges, or at which it develops into a technological civilization. These and other factors are put together in the Drake equation, namely $N_c = N^* f_p n_e f_l f_i f_c L$, stating the number of extraterrestrial technological civilizations (ETCs) in the galaxy at a given time, where the product $N^* f_p n_e$ yields the number of potentially life-supportive planets, product $f_l f_i f_c$ yields the fraction of planets in which intelligent technological life arises and $L$ is the lifetime of such ETCs. Unfortunately, the values of some of these terms are highly speculative and therefore estimates on $N_c$ vary from a few (even just one) to thousands [15] or even millions of them [16]; see also [7].

The Fermi paradox arises from the implicit assumption that $N_c$ is indeed large and that conditions for the emergence of ETCs have not changed significantly in the galactic timescale (we will return to this latter consideration in §4; check also [10]). In this scenario different answers have been suggested as a way of resolving the paradox (see e.g., [34]). While some of them rely on unverifiable conjectures about alien civilizations, many of them are based on more solid ground, such as physical considerations regarding the emergence of ETCs, or temporal issues regarding contact time. This work will initially focus on the latter class, which broadly speaking states that the time required to establish physical contact is too large, given the size of the galaxy. In this sense, although some early works [21,23] proposed that a colonization wavefront could sweep the galaxy in less than 50 Myr using ships moving at $v$=0.1c, a later analysis by Newman and Sagan [27] suggested that an ETC practicing zero population growth (a more conservative hypothesis based on the carrying capacity of the planetary environments colonized) needed a lifetime of 13 Gyr for its colonization wavefront to reach Earth.

A timescale such as that mentioned is large enough to justify the absence of contact with a colonization wavefront (even more so considering that an ETC's expansive thrust need not be constant throughout its history; see [20]) but it must be noted that other forms of contact exist, e.g., communication signals, exploration probes, etc. Regarding the case of exploration probes specifically, it is clear that they can advance much faster than a colonization wavefront, even without assuming models of self-replicating probes [6,17,33]. A very interesting recent work on this topic has been done by Bjørk [3], analyzing the timescale of galactic exploration by 4 and 8 probes, each of them endowed with a number of smaller subprobes used to explore a region around the host probe. He shows that this way of exploring the galaxy is extremely slow, being about 292 Myr required to explore only 4% of the galaxy with 8 probes. Again, this timescale may be large enough to make the chances of Earth being contacted by exploration probes from a single ETC very low. Nevertheless how these chances are affected by the fact that there might be multiple ETCs exploring the galaxy is another issue. This problem is explored here using computational simulations to obtain a model of the exploration dynamics and by doing a probabilistic analysis of the so-obtained data. The aim is to derive bounds on the number of ETCs that may be exploring the galaxy yet remain out of touch with Earth, using different assumptions on how the number of such probes is distributed and what their lifetime and their behavior is (i.e., whether they left some long-term imprint on explored systems or not).

## 2. Methodology

Following [3], the simulation of the exploration of the galaxy uses a two-level approach. To be precise, it is assumed that the exploration is carried out by space probes, each of which is capable of deploying a swarm of smaller subprobes to explore a small sector of the galaxy. Upon completing this exploration, these subprobes return to their host probe, which then proceeds to travel to another sector.

### 2.1. Simulation Model

Some uncertainty exists as to the extent of the GHZ, which might even encompass the whole galaxy [28]. In this work a GHZ encompassing stellar systems located at a radial distance ranging from 3 kpc to 11 kpc from the galactic center (GC) [26] and up to 300 pc above or below the galactic plane (GP) has been considered. This thin disk is assumed to contain $N_h=1.17 \cdot 10^{10}$ habitable stars [3], whose density $\delta_h$ decreases exponentially with the distance $r$ from the GC (with a characteristic scale length of 3 kpc), as well as with the distance $z$ above/below the GP (with a scale length of 300 pc).

The GHZ is partitioned into annular sectors spanning an angle $\alpha=\pi/600$ and whose radial length is calculated so as to comprise $N=40,000$ stars (the same value as in [3]). Notice that these sectors have a larger length in outer regions of the GHZ due to the density decline for increasing distances from the GC. The chosen value of $\alpha$ results in sectors whose width is very similar to their length (an aspect ratio between 0.89 and 1.38). Having done this, sectors placed from 3 kpc up to 11 kpc of the GC, in steps of 1 kpc are considered. For each of these nine sectors, 10 different instances are generated by sampling the density distribution given by $\delta_h$. Simulations are done on each of these instances (§2.1.1) and the results are averaged in order to obtain more meaningful estimates of the time required to explore a sector located at the corresponding distance from the GC. Subsequently, simulations are done at the galaxy level, which is modeled as a grid of sectors. These simulations are performed using a different number of probes and different radial distances from the GC for the home system from which the probes are sent (§2.1.2). Again, these experiments are replicated 10 times per parameter set in order to obtain a more representative depiction of the dynamics of the exploration.

## *2.1.1. Exploring a Sector*

The exploration of a sector is done by assuming each subprobe is assigned a non-overlapping subset of the stars in the sector. These will be visited by the subprobe (a constant velocity $v=0.1c$ is assumed), which will return to the host probe (located at the center of the sector) afterwards. The total exploration time is thus determined by the precise time the last subprobe docks back into the host probe. Clearly, both the assignment of stars to subprobes and the order in which the latter visits them have a deep impact on the total exploration time. Indeed, it can be seen that the underlying problem being faced here is the well-known vehicle routing problem [18,32]. Being an NP-hard problem, typical modern approaches for solving large instances of the problem rely on heuristic methods and most notably on meta-heuristic approaches [5].

Given the scale of the problem (tens of thousands of stars), simpler heuristic approaches have been chosen in this work. Regarding the assignment of stars to subprobes, different strategies have been considered: GP-parallel slices of a constant height, GP-parallel slices with a constant number of stars, and angular regions with a constant number of stars. The results obtained do not differ in a great extent (they are comparable within a factor of 2) and indicate that GP-parallel slices with a constant number of stars (cf. [3]) perform well. Hence this assignment procedure is considered henceforth. Once the assignment of stars to subprobes is done, the problem is reduced to several –as many as there are subprobes– instances of the conspicuous traveling salesman problem, another NP-hard problem. It is reasonable to think that if the technology for attempting large-scale exploration by space probes is available, so too will be the computational resources required to solve to optimality this kind of problems. Actually, solving instances of several tens of thousands of cities is within reach of current state-of-the-art techniques (see, e.g., [2]); even more so considering that even years of computation would be negligible with respect to the time scale of exploring a sector.

In order to get as close as possible to the optimal solution without resorting to intensive computation approaches, the nearest-neighbor heuristic (NNH) [29,3] –i.e., always picking the nearest non-yet-visited star– has been used, followed by two local improvement procedures ($\ell$-opt and 2-opt) aimed at alleviating the myopic behavior of the greedy NNH strategy (NNH solutions can get stuck with some long edges as the procedure advances). The first improvement strategy iteratively attempts to re-assign the last star visited by a subprobe to another one, if this results in an overall reduction in the exploration time. As to 2-opt [14], it is a well-known procedure that iteratively removes two non-adjacent edges from the tour and reconnects it, keeping the modification if it results in tour length reduction. The procedure is repeated until no further improvement can be found; thus the resulting tour is locally optimal with respect to any replacement of two edges (hence the name of the procedure). The combined use of these two heuristics results in a mean reduction of slightly above 10% in the overall exploration time (i.e., the time of the last subprobe).

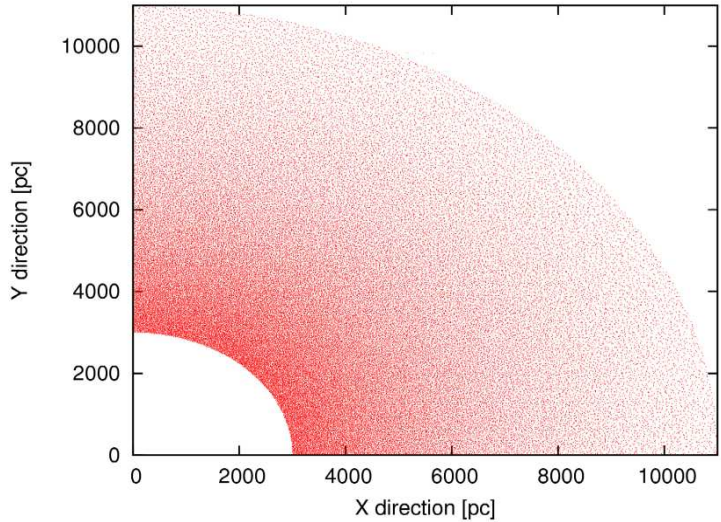

**Figure 1.** Example of a galaxy quadrant with 73,200 sectors used in the simulations.

### 2.1.2. Exploring the Galaxy

The data gathered during the simulation of the exploration of sectors is used to model the dependency of the exploration time $t$ on the distance $r$ from the GC. This will be done using a least-square fit to a function $t(r)$ –an exponential function of $r$; see §3. Once the exploration of a sector by a different number of subprobes is modeled, the exploration of the galaxy is approached as follows.

First of all, the galaxy is modeled as a grid of sectors. As in [3] a galaxy quadrant is considered (given the time scale of the simulations –50 Myr, see §3– this constitutes a large enough playground), but unlike that work, no stochastic sampling of the density function is performed to place sectors. On the contrary, a 300×244 sector grid is assumed, obtained by slicing the galaxy quadrant radially in intervals of $\alpha=\pi/600$ rad and concentrically outwards in intervals dictated by the corresponding length of sectors with $N$ stars. The coordinates of each sector are then randomly sampled within each of the boxes in this grid. This procedure has two advantages: (1) it fits the global stellar density (see Fig. 1), without being subject to large stochastic fluctuations and (2) it paves the way for analyzing the expansion of the exploration front using grid coordinates.

The exploration is performed similarly to the case of sectors. Simulations are done from different starting points placed at different distances from the GC. To be precise, 244 values –a number dictated by the division of the GHZ in sectors with $N$ stars– of $r$ are considered, spaced according to the length of sectors in the grid: let $l(r)$ be the radial length of a sector comprising $N$ stars at distance $r$ from the GC. Starting points at distances $r_1,…,r_{244}$ –given by $r_1 = 3$ kpc and $r_{j+1} = r_j + l(r_j)$– are considered. In all cases, the starting point is at an angle $\pi/4$ from the X-axis. Sectors are distributed among probes using angular slices: taking any starting reference line, an angle is chosen so that the angular slice whose vertex is in the home system contains $N_s/k$ sectors, where $N_s$ is the total number of sectors ($N_s=300\times244=73{,}200$) and $k$ is the number of probes. The first probe is assigned this slice and the procedure is repeated for the next probe, computing the angle from the boundary of the previous slice, cf. [3]. The simulation then proceeds using the NNH heuristic as in sector exploration, but including not only inter-sector travel time but also sector-exploration time (computed using the regression model calculated before). No post-optimization is performed, since the time scale of the simulation does not allow an exhaustive exploration of the quadrant. The whole procedure is repeated 10 times for each starting point and each number of probes considered, using a different quadrant generated in the same way.

## 2.2. Simulation Analysis

As mentioned before, generating the galaxy quadrant as described allows the use of grid coordinates to model the evolution of the exploration. This is particularly convenient given the symmetry of the search scenario: if probes originating in sector $(i, j)$ –which corresponds to polar coordinates $(\alpha \cdot i, r_j)$, with $\alpha$ and $r_j$ defined in the previous section– can reach sector $(i', j')$ with probability $p$ after time $t$, then the same probability exists for $(i'', j')$, where $i'' = 2i − i'$ (i.e., mirror symmetry). Furthermore, the same holds for sector $(i+\gamma, j)$ with respect to sector $(i'+\gamma, j')$, for integer $\gamma$ (i.e., circular symmetry). Since we are specifically interested in estimating the probability of probes reaching Earth, let us focus on the time at which the simulations indicate that sectors with grid coordinates $(i, j_\oplus)$ are visited, where $j_\oplus$ is the radial grid coordinate of Earth (~8,000 pc, which corresponds to $j_\oplus=191$ in the grid). More precisely, let us define $i_\oplus=150$ (without loss of generality). Now, let us consider a particular simulation in which probes originating at grid coordinates $(i, j)$ complete the exploration of sector $(i', j_\oplus)$ at time $t$. Due to the aforementioned symmetry, this is equivalent to $(i_\oplus, j_\oplus)$ being explored at time $t$ by probes originating at $(i \pm \psi, j)$, where $\psi = |i' - i_\oplus|$.

Since multiple simulations have been done, the probability of $(i_\oplus, j_\oplus)$ being explored at time $t$ by probes emitted from sector $(i, j)$ can be estimated as relative frequency of this event. To do so, given that time is a continuous variable, it is discretized in intervals of 1 Myr. As a final consideration, the analysis also includes the effect of galactic rotation, which causes a continuous flux of stars through sectors. As a first approximation the rotation curve of stars within the GHZ can be considered flat [4]. Note that due to this differential rotation, the relative position in grid coordinates of the home system from which probes were sent $t$ Myrs ago will be different at the time the probes reach Earth. The analysis centers on sector coordinates at launch time though, so we can focus on the orbital velocity of stars at radial grid coordinate $j_\oplus$. These are assumed to orbit around the GC at $2.2 \cdot 10^5$ ms$^{-1}$. Now let probes from sector $(i, j)$ start exploring sector $(i', j_\oplus)$ at time $t_1$, finishing the exploration at time $t_2$. The corresponding probability is evenly split on sectors $(i \pm \psi + \varpi, j)$, $\omega(t_1) \leq \varpi \leq \omega(t_2)$, where $\omega(t)$ is the displacement in grid coordinates of stars in sectors $(i, j_\oplus)$ after time $t$. Note that the choice of a positive sign for $\varpi$ is arbitrary, since it depends on how the positive direction of grid coordinates is interpreted.

Let $p_{ijk}(t)$ thus be the probability of a probe sent from sector $(i, j)$ reaching Earth exactly $t$ Myrs after its launch, provided that a total number of $k>0$ probes were sent. Given a collection of distributions $p_{ijk}(t)$ for different values of $k$, an aggregated probability distribution $p_{ij}(t)$ can be obtained as a weighted combination $p_{ij}(t)=\Sigma_k p(k|k>0)p_{ijk}(t)$, where $p(k|k>0)$ is the probability that the emitting civilization will use $k$ probes, given that $k>0$. The value $p^*=p(k>0)$ is the probability that in the time frame of 1 Myr a civilization will attempt to explore the galaxy with space probes. Subsequently the product $p^* p_{ij}(t)$ is the probability that a probe originating at a given star in sector $(i, j)$ will reach Earth exactly $t$ Myrs after its launch. The value of $p^*$ is a free parameter that can be modulated to analyze the resulting change in the outcome of the model (see next section). As to values $p(k|k>0)$, two scenarios are considered: $p(k|k>0) \propto k^{-1}$ (i.e., inversely proportional to the number of probes; this scenario is labeled S1 henceforth) and $p(k|k>0) \propto k^{-2}$ (i.e., inversely proportional to the total number of subprobes – see §3; this is labeled S2 henceforth). The probability $P_{ij}(t)$ of any of these probes reaching Earth at least once any time up to $t$ Myr after launch can now be recursively computed as:

$$P_{ij}(t) = \begin{cases} p_{ij}(1) & t = 1 \\ P_{ij}(t-1) + (1 - P_{ij}(t-1))p_{ij}(t) & t > 1 \end{cases} \qquad (1)$$

This calculation conservatively assumes that probes from a certain sector can reach Earth more than once (i.e., probes sent in slightly different directions might visit the same star at different times due to galactic rotation). The present time is being taken as a fixed temporal endpoint, and therefore this distribution is better interpreted as stretching forward from $t$ Myr ago until the present (the symmetry of the problem allows for this, since given any exploration interval we can always assume $(i,j)$ coordinates relative to the location $(i_\oplus, j_\oplus)$ of Earth at the beginning of that interval). This is a convenient way of modeling the problem: different values of $t$ correspond to different launches (i.e., different probes with independent trajectories) and this simplifies the subsequent analysis. The probability of a probe originating anywhere in the galaxy reaching Earth within our current time frame can now be computed. Let $T$ be the lifetime of the probes (assumed constant for simplicity). Hence, we could in principle detect or be detected by probes sent at most $T$ Myr ago. More precisely, let $q_{ij}(t)$ be the probability of not being contacted by probes from sector $(i, j)$ exactly $t$ Myr after their launch. Then, the probability $q(T)$ of not being contacted in the current Myr by probes sent from any sector up to $T$ Myr ago is

$$q(T) = \prod_{i,j}\prod_{t=1}^{T} q_{ij}(t) = \prod_{i,j}\prod_{t=1}^{T}\left(1 - p^* p_{ij}(t)\right)^N \qquad (2)$$

where $N$ is the number of stellar systems in a sector, and the inner product comprises the potential launch times before the present. A similar reasoning can be conducted to obtain the probability of Earth not having been ever explored by probes of lifetime $T$ launched at any time in the last $T'$ Myr. This scenario would correspond to active probes that left some imprint on explored systems, e.g., a beacon to mark systems with complex life (this beacon need not be an active signal emitter but could be simply a passive object of clearly artificial origin). Note that the time window $T'$ in which such exploration can take place can be larger than the lifetime of individual probes, i.e., probes may last for $T$ Myr, but they could be emitted at different times from different sectors (or even from the same one) in the last $T' \geq T$ Myr. Then, we have in this case

$$Q(T,T') = \prod_{i,j}\prod_{t=1}^{T'}\left(1-p^*P_{ij}(\min(t,T))\right)^N = $$
$$= \prod_{i,j}\left[\left(1-p^*P_{ij}(T)\right)^{N\cdot(T'-T)}\prod_{t=1}^{T}\left(1-p^*P_{ij}(t)\right)^N\right] \quad (3)$$

As mentioned before, this assumes that probes may be emitted multiple times even from the same sector, which is in principle possible (albeit extremely improbable for the parameter values considered in the experiments). A different approach can be taken assuming that probe emission from a certain sector is a non-repeatable event. Then,

$$Q(T,T') = \prod_{i,j}\left[1-\sum_{t=1}^{T'}P_{ij}(\min(t,T))\tilde{p}(1-\tilde{p})^{T'-t}\right] \quad (4)$$

where $\tilde{p} \approx Np^*$ (assuming $p^* \ll N^{-1}$) is the probability of such probes being emitted for the first time from a certain sector. In practice, the numerical difference between this model and the previous one is negligible in the simulation setting considered. For notational simplicity, we will use $q$ and $Q$ in the following, when $T$ and $T'$ are indicated in the context.

## 3. Experimental Results

The first set of simulations has been done at the sector level in order to model exploration time as a function of the distance to the GC. Experiments have been done with a number of subprobes $k \in \{4,8,16,32,64,128,256\}$. For each value of $k$, mean times required to explore a sector at distance $r$ from the GC (recall §2.1) are fitted to an exponential function $t_k(r) = a_k e^{b_k r}$ (cf. [3]). The results are shown in Fig. 2 and the fitted parameters are indicated in Table 1.

**Table 1.** Parameters $a_k$ and $b_k$ obtained fitting the time that $k$ subprobes take to explore a sector at distance $r$ from the GC to an exponential function $t_k(r) = a_k e^{b_k r}$.

| | number of subprobes | | | | | | |
|---|---|---|---|---|---|---|---|
| | 4 | 8 | 16 | 32 | 64 | 128 | 256 |
| $a_k$ [yr] | 341,576 | 178,946 | 96,704 | 56,095 | 36,312 | 26,665 | 21,756 |
| $b_k$ [pc$^{-1}$] | $1.0792\cdot10^{-4}$ | $1.0721\cdot10^{-4}$ | $1.0245\cdot10^{-4}$ | $9.3902\cdot10^{-5}$ | $8.2069\cdot10^{-5}$ | $6.7733\cdot10^{-5}$ | $5.8232\cdot10^{-5}$ |

The results obtained for 4 and 8 subprobes are comparable to those reported in [3]; for a larger number of subprobes, the parameters evolve quite regularly although for $k \geq 128$ finite effects start to be significant (the time subprobes take to get to exploration slices far from the GP and return to the host probe becomes a significant part of the total exploration time). This could be mitigated by using larger sectors, but this would also imply that the galaxy-level simulation would be more coarse-grained. To avoid potential distortions in the results due to this, up to $k=64$ subprobes will be used in subsequent simulations.

Next, simulations of the exploration of a galaxy quadrant have been performed considering $k \leq 64$ probes, each of them with the same number $k$ of subprobes. For each value of $k$ the simulations are replicated 10 times for each different position of the home sector (from grid coordinate $j = 1$ up to $j = 244$; i.e., 12,200 quadrant simulations in total). Unlike the case of sectors (in which simulations were run until the whole sector was explored and the last subprobe returned to the host probe), simulations are here done for 50 Myr of simulated time.

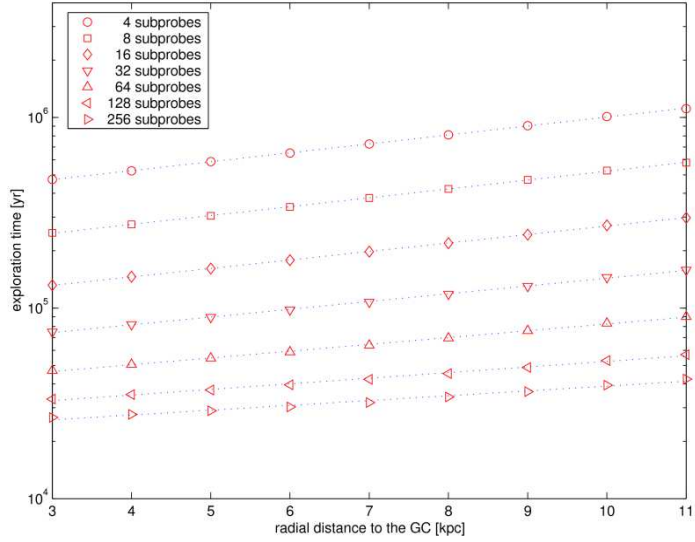

**Figure 2.** Time required to explore a sector as a function of radial distance. Error bars are negligible (standard deviation is less than 1% of the mean). In each case the dotted line is a fit to a function $t_k(r) = a_k e^{b_k r}$.

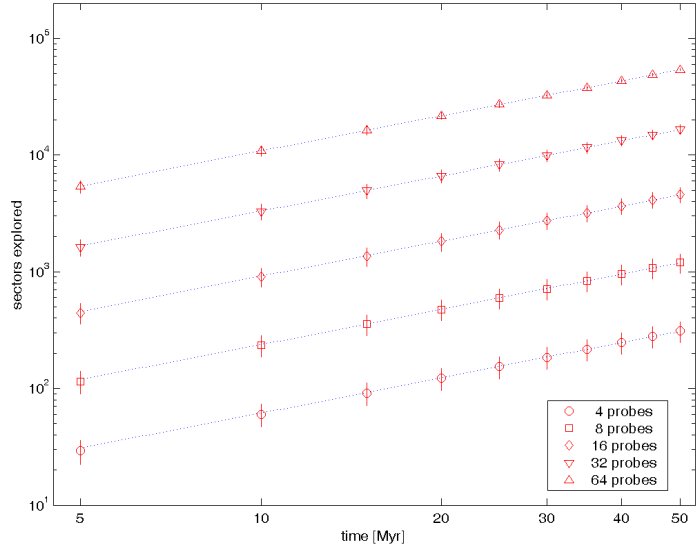

**Figure 3.** Number of sectors explored as a function of time, using different number of probes. The error bars indicate the standard deviation and each dotted line is a fit to a function $f_k(t)=\eta_k t$.

Let us firstly consider the fraction of the galaxy that can be explored in a given time. As shown in Fig. 3, the number of explored sectors grows linearly with time in all cases. A fit to a function $f_k(t)=\eta_k t$ indicates that the number of explored sectors per Myr is $\eta_4$=6.18, $\eta_8$=23.9, $\eta_{16}$=91.3, $\eta_{32}$=331 and $\eta_{64}$=1,080 for the corresponding number of probes. Thus, the projected time to explore a number of sectors equivalent to a galaxy quadrant can be estimated in this scenario (assuming this constant rate of progress in the exploration, which is optimistic and hence provides a lower bound) as 11.8 Gyr for 4 probes and 3.1 Gyr for 8 probes, a completely unrealistic time in agreement with [3]. Note however that this time decreases down to 221 Myr for 32 probes and 68 Myr for

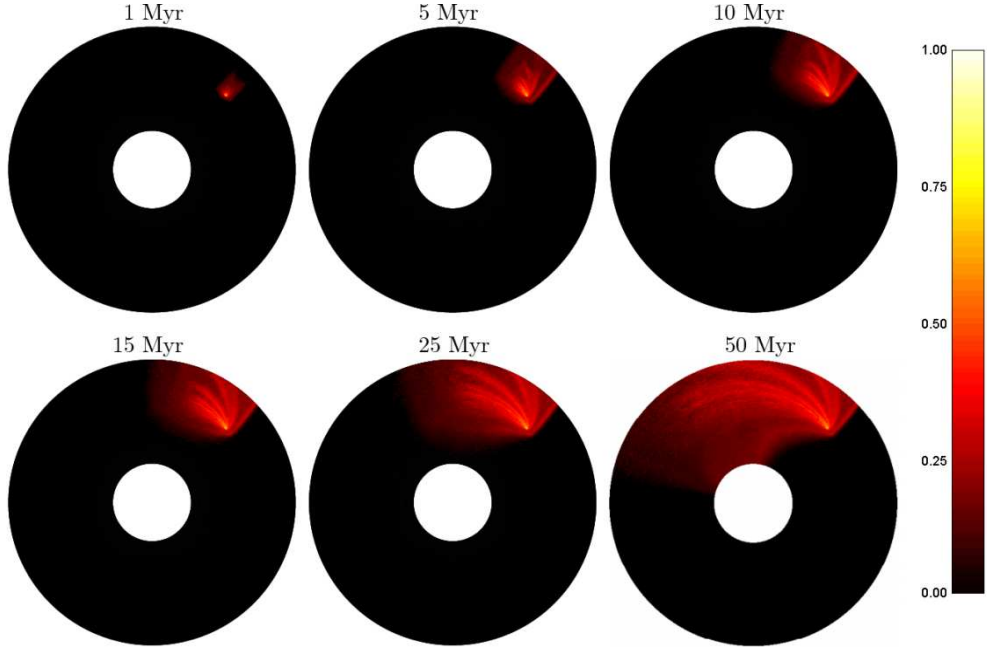

**Figure 4.** Depiction of the *hot* zone of the GHZ from which probes can reach Earth up to *T* Myr after launch. By convention, Earth is always located at an angle π/4 from the *X*-axis at the beginning of the exploration interval (*T* Myr ago) and galactic rotation is counterclockwise.

64 probes (as an aside note, some experiments with 128 and 256 probes using the parameters indicated in Table 1 indicate that the quadrant exploration time can be ~26-34 Myr and ~10-13 Myr respectively). Although these times are still high, they are not off the scale anymore. This is particularly interesting from the point of view of multiple civilizations sending exploring probes. While the chances of probes sent from a single source reaching a particular destination may be low, multiple emissions from different sources would substantially increase these chances.

Using the data from the simulations the distributions $p_{ij}(t)$ and $P_{ij}(t)$ have been computed as described in §2.2. Fig. 4 illustrates the latter distribution. As expected, the distribution is elongated due to galactic rotation and is denser near Earth and in outer regions of the GHZ, since the higher concentration of stars towards the GC attracts more probes to sectors located at lower *r* coordinate. These distributions are used to compute ***q*** and ***Q***, the probability of probes *not* visiting Earth within the last Myr or the last *T'* Myr respectively. As mentioned before, a crucial parameter to compute these values is $p^*$, the probability of probes being sent from a certain star in a certain Myr. Given the enormous number of potentially habitable systems stellar systems in the GHZ, a large enough value of $p^*$ can make it highly unlikely that no probe has ever reached Earth. This is depicted in Fig. 5, in which both ***q*** and ***Q*** –considering a time window *T'* = 100 Myr, an upper estimate of the Fermi-Hart timescale [21,31,12]– are represented as a function of $p^*$ in scenarios S1 and S2. As can be seen, there is a region of about two orders of magnitude in $p^*$ making ***Q*** change from near 0 (highly unlikely that probes have not visited Earth) to near 1 (highly likely that probes have not visited Earth). The same holds for ***q*** both in S1 and S2.

Assuming Earth has not been visited in the last 100 Myr (something that can be stated referring to sterilizing probes, or to probes leaving perdurable beacons behind), this event is statistically unlikely (using ***Q***<0.05 as a proxy for this) if $p^*>3.77 \cdot 10^{-10}$ [S1] ($p^*>1.35 \cdot 10^{-9}$ [S2]) if the lifetime of probes is 50 Myr. Given $N_h=1.17 \cdot 10^{10}$ habitable stars in the GHZ, this translates to $N_h p^* \sim$ 4-16 civilizations at most sending such probes in any given Myr (this probability is assumed constant throughout the whole GHZ). In a more limited scenario in which contact is restricted to the last Myr, this cutoff probability is $p^*$=2.69·10$^{-8}$ [S1] ($p^*$=9.93·10$^{-8}$ [S2]), resulting in approximately 315-1,160 civilizations exploring the galaxy in a given Myr. These values depend obviously on the lifetime of probes. Fig. 6 shows how the cutoff value of $p^*$ varies along with the lifetime $T$ of probes. As it can be seen, the relationship between these values follows a power-law $p^* \propto T^{-\gamma}$, where $\gamma \approx 1$ (a least-squares fit indicates values between 1.008 [$Q$, S1] and 1.086 [$q$, S2]). This relationship seems to be robust with respect to the time window and the particular distribution of number of probes considered and could be used to project the critical value of $p^*$ for larger probe lifetimes.

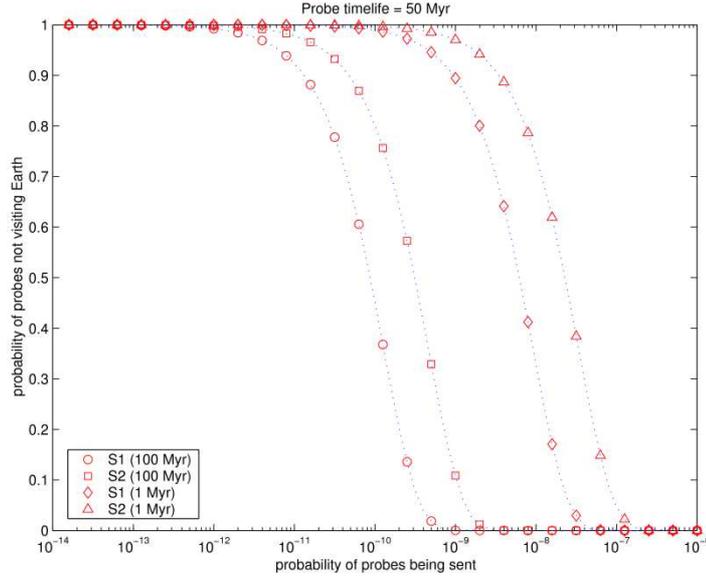

**Figure 5.** Probability of probes not visiting Earth as a function of $p^*$, the probability of probes being sent from a certain star at a certain Myr. Probe lifetime is $T$=50 Myr and the time window $T'$ for computing $Q$ is 100 Myr.

## 4. Discussion

The results of the simulations indicate that the timescale for exploring the galaxy with non-replicating probes is certainly very high if a reduced number of probes is used (~12 Gyr for 4 probes and ~3 Gyr for 8 probes, just for exploring 25% of the galaxy). The question of whether an ETC might choose to attempt such an exploration in light of the enormous time required can however be ultimately answered only in speculative terms, since this would imply making unjustified assumptions about the motivations and goals of a completely unknown species (it should be noted that automated probes may keep on exploring long after their creators cease to exist). At any rate, this issue is independent of the analysis conducted, which has considered a small time window in galactic terms, consistent for example with local exploration initiatives. The situation is somehow different if a large number of probes is used (~40-60 Myr for 256 probes with 256 subprobes each). This latter timescale falls within the exploration range of self-replicating probes, suggested by Tipler [31] to be ~300 Myr (actually an overoptimistic bound, due to the implied mass consumption rate as pointed out by Sagan and Newman [30]). This may have an impact on the relative risk/benefit trade-off of this latter kind of probes, since they could be dominated in a Pareto sense (given comparable exploration times, the less risky option is preferable) by an approach based on a large number of non-replicating probes, with the implications that this would have for an optimization-oriented ETC [11].

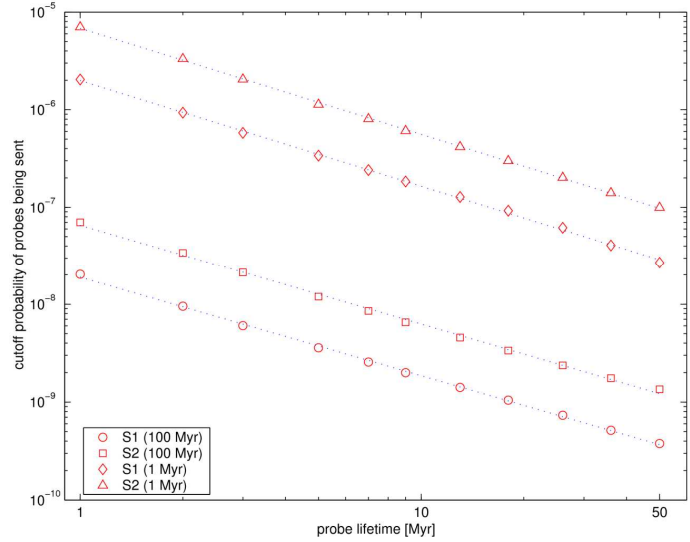

**Figure 6.** Relationship between the cutoff probability of probes being emitted and the lifetime of probes. The dotted lines represent a fit to a function $p(t) = at^b$.

Regarding the critical number of ETCs sending out exploration probes above which non-contact becomes statistically unlikely, different estimates have been found depending on whether contact close to our current time, or at any time during the last $T'$ Myr is considered. In the first case, this number may range from a few hundreds to several thousands per Myr. A Monte-Carlo simulation of the spatial distribution of these ETCs (using the density function $\delta_h$) indicates that the nearest one would be at an average distance of ~150 pc for 10,000 ETCs, ~400 pc for 1,000 ETCs, ~1,100 pc for 100 ETCs and ~3,200 pc for 10 ETCs in the GHZ. These distances are large enough, at least in their upper range, to make difficult the detection of a Kardashev Type II ETC that was not directly signaling at us and even in this case detection via traditional SETI approaches would be hard due to phenomena such as e.g., interstellar scintillation [13] (these inherent limitations strengthen the case for alternatives strategies other than radio-listening in the water hole region; see for example [22]). This is even more so in a scenario as suggested in [11], in which these advanced ETCs follow an optimization approach as their existential imperative and hence would try to avoid any energy leakage. Failure to detect Dyson spheres via their infrared signature [24,8] also supports this (a recent report [9] indicates that a search on the IRAS Low Resolution Spectrometer database sensitive enough to find solar-sized Dyson spheres out to 300 pc left only 4 moderately interesting yet still ambiguous candidates). It is suggested [11] that these introverted ETCs would use small research probes that would act as some kind of distributed sensorial apparatus. The bounds obtained in this work are not small enough to provide conclusive support of the unlikeliness of ETCs using intrusive probes in a short timescale though.

The implications of absence of contact with perdurable beacons left behind by exploring probes are also interesting. Independently of the reasons (which once again can only be conjectured) why an ETC might choose to attempt this kind of endeavor or not, it must be noted than the longer the time window for leaving such beacons, the smaller the corresponding critical value of $p^*$ becomes (e.g., $p^*$=2.41·10$^{-11}$ for $T'$ = 1 Gyr and $p^*$=7.92·10$^{-12}$ for $T'$ = 3 Gyr [S1, $T$ = 50 Myr]). This would demand a robust explanation of why no ETC attempted it if indeed there were many spacefaring ETCs at any time during galactic history. This problem can be solved if a physical mechanism poses a strict limit on the size of this time window. Such a mechanism was proposed by Annis [1], introducing an astrobiological disequilibrium hypothesis. In this model, gamma-ray bursts act as a regulation mechanism, constituting a source of global catastrophic events that wipe out most complex life in the GHZ. This model has been further studied by Ćirković and Vukotić [12], suggesting that we are now immersed in phase transition between a regime in which there are numerous astrobiological resetting events and a regime in which the natural decline in such events will result in a suitable environment for the emergence of numerous ETCs, eventually leading to the one or more Kardashev Type III ETCs. The duration of this phase transition is proposed to be the Fermi-Hart timescale. The outcome of the analysis at this timescale is compatible with this hypothesis: a GHZ mostly devoid of complex life ~100 Myr ago and in which ETCs are gradually appearing does not require any additional assumption to account for a low maximum number of ETCs performing the kind of exploration considered.

## 5. Conclusions

Exploring the galaxy with space probes would take a long time unless a high number of probes were used. A large enough number of ETCs conducting exploration initiatives can however make it very unlikely that no probe ever reach a particular stellar system. According to the numerical model devised, the number of ETCs actively exploring the galaxy per Myr is bounded from above by ~$10^2$-$10^3$ (assuming probes have a lifetime of 50 Myr –which can be reasonably thought to fall within the technological reach of a Kardashev Type II ETC and can be considered a short lapse in the galactic timescale– and that contact evidence lasts for 1 Myr). In this scenario, the nearest such ETC would be on average from several hundred up to around one thousand parsecs distance from Earth. This bound sharply decreases for fixed probe lifetime as contact evidence becomes longer-lasting (e.g., ~10 for 100 Myr), as it would be the case with, e.g., sterilizing probes. A parsimonious non-exclusive explanation to this diminishing upper bound can be found in non-stationary astrobiological models, which impose a strict temporal limit for the emergence of complex life and hence in the time window for contact. Future work will try to confirm these conclusions on different simulation models, including a sensitivity analysis of the parameters involved.

## Acknowledgements

We would like to thank the reviewers for their constructive comments and suggestions. Thanks are also due to José Muñoz for useful discussions and to Spanish MICINN for support under project TIN2008-05941.

## References

[1] J. Annis. "An astrophysical explanation for the Great Silence", *JBIS*, **52**:19–22, 1999.
[2] D.L. Applegate, R.E. Bixby, V. Chvátal, W. Cook, D.G. Espinoza, M. Goycoolea, and K. Helsgaun. "Certification of an optimal TSP tour through 85,900 cities", *Operations Research Letters*, **37**:11–15, 2009.
[3] R. Bjørk. "Exploring the galaxy using space probes", *International Journal of*

*Astrobiology*, **6**:89–93, 2007.

[4] A. Bosma. "21-cm line studies of spiral galaxies. II. The distribution and kinematics of neutral hydrogen in spiral galaxies of various morphological types", *Astronomical Journal*, **86**:1825-1846, 1981

[5] C. Blum and A. Roli. "Metaheuristics in combinatorial optimization: Overview and conceptual comparison", *ACM Computing Surveys*, **35**(3):268–308, 2003.

[6] R.N. Bracewell. "Communications from superior galactic communities", *Nature*, **186**:670–671, 1960.

[7] M.J. Burchell. "W(h)ither the drake equation?" *International Journal of Astrobiology*, **5**(3):243–250, 2006.

[8] R.A. Carrigan. "Searching for Dyson spheres with Planck spectrum fits to IRAS", In 55*th International Astronautical Congress*, Vancouver, Canada, 2004. IAC-04-IAA-11106.

[9] R.A. Carrigan. "IRAS-based whole-sky upper limit on Dyson spheres", Technical Report 08-352, Fermilab, 2008. astro-ph:0811-2376.

[10] M.M. Ćirković. "The temporal aspect of Drake equation and SETI", *Astrobiology*, **4**:225–231, 2004.

[11] M.M. Ćirković. "Against the empire", *JBIS*, **61**(7):246–254, July 2008.

[12] M.M. Ćirković and B. Vukotić. "Astrobiological phase transition: Towards resolution of Fermi's paradox", *Origins of Life and Evolution of Biospheres*, **38**:535–547, 2008.

[13] J.M Cordes, T.J.W. Lazio and C. Sagan. "Scintillation-induced intermittency in SETI", *Astrophysical Journal*, **487**:782-808, 1997

[14] G.A. Croes. "A method for solving traveling salesman problems", *Operations Research*, **6**:791–812, 1958.

[15] M.J. Fogg. "Temporal aspects of the interaction among the first galactic civilizations: The «interdict hypothesis»", *Icarus*, **69**(2):370–384, 1987.

[16] J. Freeman and M. Lampton. "Interstellar archaeology and the prevalence of intelligence", *Icarus*, **25**(2):368–369, 1975.

[17] R.A. Freitas Jr. "A self-reproducing interstellar probe", *JBIS*, **33**:251–264, 1980.

[18] B.L. Golden and A.A. Assad. "*Vehicle Routing: Methods and Studies*, **vol. 16** of *Studies in Management Science and Systems*", North-Holland, Amsterdam, 1988.

[19] G. Gonzalez, D. Brownlee, and P. Ward. "The galactic habitable zone: Galactic chemical evolution", *Icarus*, **152**(1):185–200, 2001.

[20] C. Gros. "Expanding advanced civilizations in the universe", *JBIS*, **58**(2):108–111, 2005.

[21] M.H. Hart. "An explanation for the absence of extraterrestrials on Earth", *Quarterly Journal of the Royal Astronomical Society*, **16**:128–135, 1975.

[22] A.W. Howard *et al.* "Search for nanosecond optical pulses from nearby solar-type stars", *Astrophysical Journal*, **613**:1270-1284, 2004

[23] E.M. Jones. "Colonization of the galaxy", *Icarus*, **28**(3):421–422, 1976.

[24] J. Jugaku and S.E. Nishimura. "A search for Dyson spheres around late-type stars in the solar neighborhood", In R. Norris and F. Stootman (Eds.), *Bioastronomy 2002: Life Among the Stars*, **vol. 213** of *Proceedings of the IAU Symposium*, page 437, San Francisco CA, 2002. Astronomical Society of the Pacific.

[25] C.H. Lineweaver. "An estimate of the age distribution of terrestrial planets in the universe: Quantifying metallicity as a selection effect", *Icarus*, **151**(2):307–313, 2001.

[26] C.H. Lineweaver, Y. Fenner, and B.K. Gibson. "The galactical habitable zone and the age distribution of complex life in the Milky Way", *Science*, **303**:59–62, 2004.

[27] W.I. Newman and C. Sagan. "Galactic civilizations: Population dynamics and interstellar diffusion", *Icarus*, **46**(3):293–327, 1981.

[28] N. Prantzos. "On the «galactic habitable zone»", *Space Science Reviews*, **135**(1–4):313–322, 2008.

[29] G. Reinelt. "*The Traveling Salesman Problem: Computational solutions for TSP applications*", **vol. 840** of *Lecture Notes in Computer Science*. Springer-Verlag, Berlin Heidelberg, 1994.

[30] C. Sagan and W.I. Newman. "The Solipsist Approach to Extraterrestrial Intelligence", *Quarterly Journal of the Royal Astronomical Society*, **24**:113-121, 1983

[31] F.J. Tipler. "Extraterrestrial intelligent beings do not exist", *Quarterly Journal of the Royal Astronomical Society*, **21**:267–281, 1980.

[32] P. Toth and D. Vigo. "*The Vehicle Routing Problem*", **vol. 9** of *Monographs on Discrete Mathematics and Applications*. SIAM, Philadelphia PN, 2001.

[33] F. Valdes and R.A. Freitas Jr. "Comparison of reproducing and nonreproducing starprobe strategies for galactic exploration", *JBIS*, **33**:402–406, 1980.

[34] S. Webb. "*If the Universe is teeming with aliens... where is everybody? Fifty solutions to the Fermi paradox*", Springer, New York NY, 2002.